# Sea Quark Effects on the Strong Coupling Constant *

T. Onogi[a†], S. Aoki[b], M. Fukugita[c], S. Hashimoto[a], N. Ishizuka[b], H. Mino[d], M.Okawa[e], A. Ukawa[b]

[a]Department of Physics, Hiroshima University, Higashi-Hiroshima 724, Japan

[b]Institute of Physics, University of Tsukuba, Ibaraki 305, Japan

[c]Yukawa Institute for Theoretical Physics, Kyoto University , Kyoto 606, Japan

[d]Faculty of Engineering, Yamanashi University, Kofu 400, Japan

[e]National Laboratory for High energy Physics (KEK), Ibaraki 305, Japan

We present results showing that the strong coupling constant measured in two-flavor full QCD with dynamical Kogut-Susskind quarks at $\beta = 5.7$ exhibit a 15% increase due to sea quarks over that for quenched QCD at the scale $\mu \approx 7\text{GeV}$ .

## 1. Introduction

Recently a lattice QCD determination of the strong coupling constant $\alpha_s$ has been attempted employing the spin-averaged charmonium $1S-1P$ mass splitting to fix the scale[1]. The calculations are so far restricted to quenched QCD, and corrections due to sea quarks have to be estimated to make a physical prediction for $\alpha_s$. In this article we report on our calculation of the strong coupling constant in two-flavor full QCD in the spirit of Ref. [1] and discuss implication of the results on the effects of sea quarks.

## 2. Simulation

Our study is based on the full QCD configurations on a $20^4$ lattice previously generated with two flavors of dynamical Kogut-Susskind quarks at $\beta = 5.7$ with $m_q a = 0.01$[2]. For charmonium spectrum measurement we employ a subset of these configurations, periodically doubled in the temporal direction. We use both Wilson (without 'clover' improvement) and Kogut-Susskind actions for valence quarks. The Wilson results are obtained at $K = 0.130$ and 0.135 with 72 and 75 configurations, and the preliminary Kogut-Susskind results, obtained after the conference, are based on 19 configurations at $m_q a = 0.3$ and 0.4.

For Wilson valence quarks we used a Gaussian smeared source and a local sink. Typical forms of the operators are listed in Table 1. Two types of operators are used for the ${}^1P_1$ state $h_c$, labeled $P1$ and $P2$.

For Kogut-Susskind valence quarks we employ operators constructed in Ref. [3], which are also

Table 1
Charmonium operators. Smearing functions used for the Wilson case are $f_s(x) = \exp(-|x|^2/4)$ and $f_p^i(x) = \sin(2\pi x^i/L) f_s(x)$. $\epsilon_x = (-1)^{x_1+x_2+x_3}$ and $\zeta_{xi} = \epsilon_x (-1)^{x_i}$.

| meson | source | sink |
|---|---|---|
| (a) Wilson quark | | |
| $J/\psi$ | $\sum_{x,y} \bar\psi_x \gamma^i \psi_y f_s(x) f_s(y)$ | $\bar\psi_x \gamma^i \psi_x$ |
| $\eta_c$ | $\sum_{x,y} \bar\psi_x \gamma^5 \psi_y f_s(x) f_s(y)$ | $\bar\psi_x \gamma^5 \psi_x$ |
| $h_c$(P1) | $\sum_{x,y} \bar\psi_x \sigma^{jk} \psi_y f_s(x) f_s(y)$ | $\bar\psi_x \sigma^{jk} \psi_x$ |
| $h_c$(P2) | $\sum_{x,y} \bar\psi_x \gamma^5 \psi_y f_p^i(x) f_s(y)$ | $\bar\psi_x \sigma^{jk} \psi_x$ |
| (b) Kogut-Susskind quark | | |
| $J/\psi$ | $\sum_x \zeta_{xi} \bar\chi_{x,t} \chi_{x,t}$ | |
| $\eta_c(NG)$ | $\sum_x \epsilon_x \bar\chi_{x,t} \chi_{x,t}$ | |
| $\eta_c$ | $\sum_x (\bar\chi_{x,t}\chi_{x,t+1} - \bar\chi_{x,t+1}\chi_{x,t})$ | |
| $h_c$ | $\sum_x \zeta_{xi} (\bar\chi_{x,t}\chi_{x,t+1} + \bar\chi_{x,t+1}\chi_{x,t})$ | |

*presented by T. Onogi

†present address: Theory Group, Fermi National Accelerator Laboratory, Batavia, IL 60510, U. S. A.

Table 2
Charmonium $1S-1P$ splitting in full QCD.

| operator | $K/m_q a$ | $\Delta m_{1S-1P}a$ | $\pi/a$(GeV) |
|---|---|---|---|
| (a) Wilson valence quark | | | |
| $h_c(P1)$ | 0.130 | 0.221(49) | 6.5(1.5) |
| | 0.135 | 0.221(48) | 6.5(1.4) |
| $h_c(P2)$ | 0.130 | 0.200(32) | 7.2(1.2) |
| | 0.135 | 0.205(35) | 7.0(1.2) |
| (b) Kogut-Susskind valence quark | | | |
| $\eta_c(NG)$ | 0.3 | 0.185(20) | 7.79(84) |
| | 0.4 | 0.208(23) | 6.93(77) |
| $\eta_c$ | 0.3 | 0.148(20) | 9.7(1.3) |
| | 0.4 | 0.164(23) | 8.8(1.2) |

listed in Table 1. For $\eta_c$ the first operator corresponds to the Nambu-Goldstone channel of $U(1)$ chiral symmetry of the Kogut-Susskind action. For sources we employ an improved wall source technique described in Ref. [4].

## 3. Determination of scale

### 3.1. Wilson valence quarks

In Table 2(a) we list our result for the charmonium $1S-1P$ mass splitting obtained with Wilson valence quarks and the corresponding scale $\pi/a$. We observe unusually large errors of 20–25% in spite of the use of over 70 configurations. In contrast a test run in quenched QCD on a $16^3 \times 32$ lattice at $\beta = 6.0$ with 20 configurations showed only a 7% error, and the full QCD results with Kogut-Susskind valence quarks described below also exhibit smaller errors with 19 configurations. A possible cause of the large errors for Wilson valence quarks might be ascribed to the mismatch of quark actions taken for sea and valence quarks, though actual connection is not clear to us.

### 3.2. Kogut-Susskind valence quarks

It is more natural to employ the Kogut-Susskind quark action for valence quarks on our full QCD configurations generated with the same action for dynamical sea quarks. Our results for the $1S-1P$ splitting obtained for this case are listed in Table 2(b).

In Fig. 1 we show the charmonium spectrum for the valence quark mass $m_q a = 0.3$. Conversion to physical units is made with $\pi/a$ obtained with $\eta_c(NG)$ given in Table 2. Except for the degeneracy of $J/\psi$ and $\eta_c$ in the non-Nambu-Goldstone channel, the pattern of the spectrum is consistent with the experiment. In particular $\eta_c$ in the Nambu-Goldstone channel is lighter than $J/\psi$ (the value of the mass splitting $m_{J/\psi} - m_{\eta_c} = 360(40)$MeV is three times larger than the experimental value 118MeV, however). This is quite different from the results with the Wilson action for which the addition of the clover term is needed to split the two $1S$ states. We note that the errors in the physical values of masses shown in Fig. 1 do not include that of the scale $\pi/a$ of 11%, which translates to about 400MeV for the masses. An increase in statistics is needed to see if the physical charm quark corresponds to a smaller value of $m_q a$.

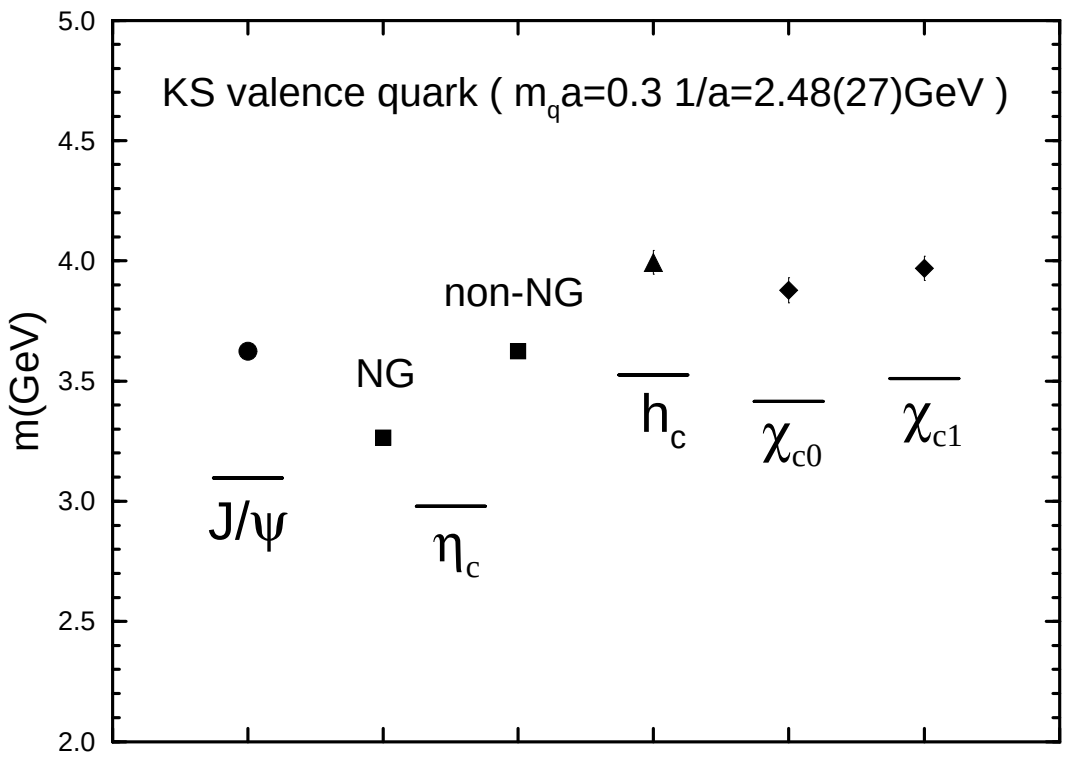

Figure 1. Charmonium spectrum in two-flavor full QCD at $\beta = 5.7$ with Kogut-Susskind sea ($m_q a = 0.01$) and valence ($m_q a = 0.3$) quarks. Errors shown do not include that of scale. Horizontal lines indicate experimental values.

## 4. Strong coupling constant

Our full QCD results for the $1S-1P$ mass splitting yield a value $\pi/a \approx 7$GeV with an error of 10–20% for Wilson and Kogut-Susskind valence quarks (the latter with the Nambu-Goldstone $\eta_c$). This value is consistent with $\pi/a = 7.01(28)$[2] estimated from the $\rho$ meson mass using Kogut-Susskind valence quarks. The use of the non-Nambu-Goldstone operator for $\eta_c$, however, leads to a value 30% larger for the scale. This discrepancy represents a systematic uncertainty for the Kogut-Susskind case.

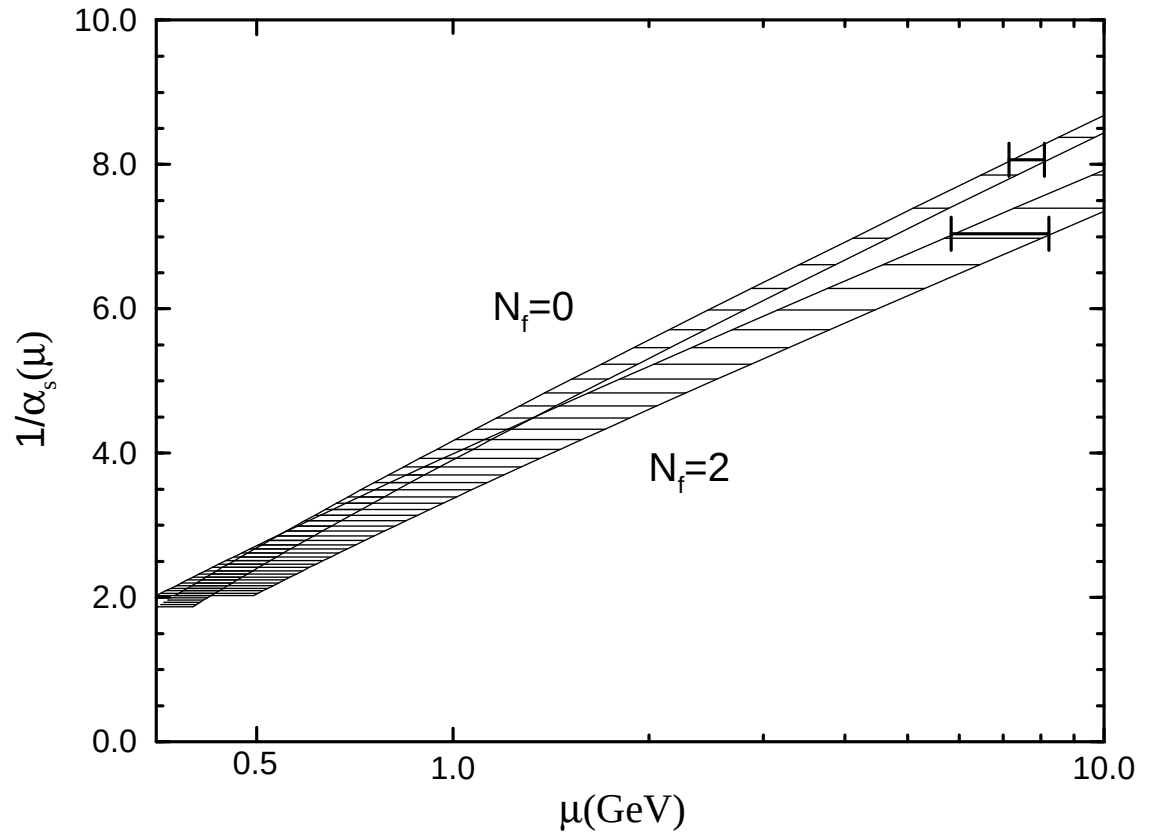

Figure 2. Evolution of $\alpha_s(\mu)$ for quenched and two-flavor QCD. Thick bars show measured values of $\alpha_s(\mu)$ and uncertainties in the scale.

In order to estimate the strong coupling constant at the scale estimated above, we employ the relation[1,5],

$$\alpha_{\overline{MS}}^{(N_f)}(\pi/a)^{-1} = P\alpha_0^{-1} + 0.3093 - 0.0885N_f, \quad (1)$$

where the last term represents the contribution of $N_f$ flavors of Kogut-Susskind sea quarks[6]. With $P = 0.577323(34)$[2] at $\beta = 5.7$ and $N_f = 2$ this gives

$$\alpha_{\overline{MS}}^{(2)} = 0.142. \quad (2)$$

For comparison we quote the quenched result:

$$\alpha_{\overline{MS}}^{(0)} = 0.124, \quad (3)$$

for $\pi/a = 7.63(48)$GeV at $\beta = 6.1$[1]. We thus find that the strong coupling constant extracted for two-flavor full QCD is about 15% larger than in the quenched case at $\mu \approx 7$GeV.

A larger value is in fact expected from the consideration that fixing the scale by the $1S - 1P$ mass splitting means adjusting the QCD coupling strength at a scale typical of charmonium and that the coupling for larger momenta decreases more slowly in the presence of sea quarks. This view underlies the heuristic procedure of Ref. [1] in which the value of $\alpha_{\overline{MS}}^{(4)}$ was estimated from the measured value of $\alpha_{\overline{MS}}^{(0)}$ by matching the two couplings evolved down to the scale $\mu \approx 0.4 - 0.75$GeV with the two-loop renormalization group $\beta$ function.

In Fig. 2 we apply this procedure to $\alpha_{\overline{MS}}^{(2)}$ and $\alpha_{\overline{MS}}^{(0)}$ starting with (2) (the Wilson result with $h_c(P2)$ at $K = 0.135$ is used for $\pi/a$; other cases are similar) and (3). The overlap of the two bands at $\mu \approx 0.5$GeV shows that a 15% increase found for $\alpha_{\overline{MS}}^{(2)}$ is quantitatively consistent with the expected magnitude of sea quark effects. This agreement also implies that our two-flavor result would lead to a value of $\alpha_{\overline{MS}}^{(4)}$ similar to the original result $\alpha_{\overline{MS}}^{(4)}(5\text{GeV}) = 0.172(12)$[1]. In fact we find $\alpha_{\overline{MS}}^{(4)}(5\text{GeV}) = 0.172^{+0.012}_{-0.009}$ using our Wilson result with $h_c(P2)$ at $K = 0.135$ in Table 2.

## 5. Summary

Our results and analyses show that sea quark effects are visible in the strong coupling constant measured in current full QCD simulations. This indicates a promising prospect in the near future for a full QCD determination of the strong coupling constant including a more realistic spectrum of sea quarks than was attempted here.

## Acknowledgement

We thank A. El-Khadra, A. Kronfeld and P. Mackenzie for informative discussions.

## REFERENCES

1. A. X. El-Khadra *et.al.*, Phys. Rev. Lett. **69** (1992) 729.
2. M. Fukugita *et.al.*, Phys. Rev. D**47** (1993) 4739.
3. M. F. L. Golterman, Nucl. Phys. **B273** (1986) 663.
4. N. Ishizuka *et.al.*, preprint KEK-TH-351 (1993) (Nucl. Phys. B, in press).
5. G. P. Lepage and P. Mackenzie, Phys. Rev. D**48** (1993) 2250.
6. H. S. Sharatchandra *et.al.*, Nucl. Phys. **B192** (1981) 205.